\documentclass[pre,twocolumn,superscriptaddress]{revtex4}

\usepackage{amssymb,amsbsy}
\usepackage{graphicx}
\usepackage{rotating}
\usepackage{dcolumn}
\usepackage{amsmath}
\usepackage[abs]{overpic}
\usepackage{xr}
\usepackage{xcolor}

\newcommand{\abs}[1]{\vert #1 \vert}
\newcommand{\av}[1]{\langle #1 \rangle}

\newcommand{\etal}{\textit{et al.}}
\newcommand{\mat}{\textbf}
\renewcommand{\vec}{\textbf}

\renewcommand{\eqref}[1]{\ref{#1}}
\newcommand{\citei}[1]{{\cite{#1}}}

\begin{document}
\title{Belief propagation for networks with loops}

\author{Alec Kirkley}
\affiliation{Department of Physics, University of Michigan, Ann Arbor, Michigan 48109, USA}

\author{George T. Cantwell}
\affiliation{Department of Physics, University of Michigan, Ann Arbor, Michigan 48109, USA}
\affiliation{Santa Fe Institute, 1399 Hyde Park Road, Santa Fe, NM 87501, USA}

\author{M. E. J. Newman}
\affiliation{Department of Physics, University of Michigan, Ann Arbor, Michigan 48109, USA}
\affiliation{Santa Fe Institute, 1399 Hyde Park Road, Santa Fe, NM 87501, USA}
\affiliation{Center for the Study of Complex Systems, University of Michigan, Ann Arbor, Michigan 48109, USA}

\begin{abstract}
  Belief propagation is a widely used message passing method for the solution of probabilistic models on networks such as epidemic models, spin models, and Bayesian graphical models, but it suffers from the serious shortcoming that it works poorly in the common case of networks that contain short loops.  Here we provide a solution to this long-standing problem, deriving a belief propagation method that allows for fast calculation of probability distributions in systems with short loops, potentially with high density, as well as giving expressions for the entropy and partition function, which are notoriously difficult quantities to compute.  Using the Ising model as an example, we show that our approach gives excellent results on both real and synthetic networks, improving significantly on standard message passing methods.  We also discuss potential applications of our method to a variety of other problems.\\
  
\end{abstract}
\maketitle 

\section{Introduction}
\label{sec:intro}
Many complex phenomena can be modeled using networks, which provide powerful abstract representations of systems in terms of their components and interactions~\citei{newman2018networks}.  Phenomena of interest are often modeled using probabilistic formulations that capture the probabilities of states of network nodes.  Examples include the spread of epidemics through networks of social contacts~\citei{kiss2017mathematics}, cascading failures in power grids~\citei{carreras2004dynamical}, and the equilibrium behavior of spin models such as the Ising model~\citei{DGM02b}.  Networks are also used to represent pairwise dependencies between variables in statistical models that do not otherwise have a network component, as a convenient tool for bookkeeping and visualization of model structure~\citei{jordan1998learning}.  Such ``graphical models,'' which allow us to represent the conditional dependencies between variables in a non-parametric manner, form the foundation for many modern machine learning techniques~\citei{koller2009probabilistic}.

The solution of probabilistic models like this presents a challenge.  Analytic methods such as those used for regular lattices do not generalize to the more complex topologies of networks, and mean-field and other standard approximations often fail to take crucial details of network structure into account.  Numerical methods can be successful but are computationally demanding on larger networks and sometimes give results of poor accuracy.  Message passing or ``belief propagation'' methods offer an alternative and promising approach that straddles the line between analytic and numerical techniques~\citei{bethe1935statistical,mezard2009information}.  Message passing works by deriving a set of self-consistent equations satisfied by the variables or probabilities of interest and then solving those equations by numerical iteration.  The name ``message passing'' comes from the fact that the equations can be thought of as representing messages passed between neighboring nodes in the network.

Standard formulations of message passing, however, have a crucial weakness: they rely on the assumption that the states of the neighbors are uncorrelated with one another, which is only true if the network contains no loops.  Unfortunately, almost all real-world networks do contain loops, and usually many of them~\citei{watts1998collective}, so standard message passing can give quite poor results in practical situations.  In this paper we propose a solution to this problem in the form of a new class of message passing methods for probabilistic models on ``loopy'' networks.  These methods open up a host of possibilities for novel network calculations, many of which we discuss here.

The limitations of traditional message passing have been widely noted in the past and a number of previous attempts have been made to remedy them.  The only truly loopless networks are trees, but standard message passing methods have been shown to give good results on networks that satisfy the weaker condition of being ``locally tree-like,'' meaning that local regions of the network take the form of trees even though the network as a whole is not a tree.  In effect, this means that the network can contain long loops, but not short ones~\citei{newman2018networks}.  However, realistic networks often fail to satisfy even this weaker condition and contain many short loops.  Message passing has been extended to certain classes of random graphs with short loops, such as Husimi graphs~\citei{EKBV05,MNB11,BMN12} and other tree-like agglomerations of small loopy subgraphs~\citei{newman2009random,yoon2011belief}, but these techniques are not generally applicable to real-world networks.  Alternatively, one can incorporate the effect of loops by using a perturbative expansion around the loopless case~\citei{montanari2005compute,chertkov2006loop}, though this approach becomes progressively less accurate as the number of loops increases and is therefore best suited to networks with a low loop density, which rules out a large fraction of real networks, whose loop density is often high~\citei{watts1998collective,newman2003social}.  Perhaps the best known extension of belief propagation, and the one most similar to our own approach, is the method known as generalized belief propagation~\citei{yedidia2001generalized}, which is based on the idea of passing messages not just between pairs of nodes but between larger groups. This method is however quite complicated to implement and requires explicit construction of the groups, which involves nontrivial pre-processing steps~\citei{welling2004choice}.  The method we propose requires no such steps.

In Ref.~\citei{cantwell2019message} we previously described message passing schemes for percolation models and spectral calculations on loopy networks.  In this paper we extend this approach to the solution of general probabilistic models.  We derive a factorization of the probability of states for such models that allows us to write self-consistent message passing equations for the marginal probabilities on sets of nodes in a neighborhood around a given reference node.  From these equations we can then calculate a range of quantities of interest such as single-site marginals, partition functions, and entropies.  To ground our discussion we use the Ising model as an example of our approach, showing how our improved message passing methods can produce better estimates for this model than regular message passing.  We show that our methods are asymptotically exact on networks whose loop structure satisfies certain general conditions and give good approximations for networks that deviate from these conditions.  We give example results for the Ising model on both real and artificial networks and also discuss applications of our method to a range of other problems, emphasizing its wide applicability.

\section{Materials and Methods}
Our first step is to develop the general theory of message passing for probabilistic models on loopy networks.  With an eye on the Ising model, our discussion will be in the language of spin models, although the methods we describe can be applied to any probabilistic model with pairwise dependencies between variables, making it suitable for a broad range of calculations in probabilistic modeling.

\subsection{Model description}
Consider a general undirected, unweighted network~$G$ composed of a set~$V$ of nodes or vertices and a set~$E$ of pairwise edges.  The network can be represented mathematically by its adjacency matrix~$\mat{A}$ with elements $A_{ij}=1$ when nodes $i$ and $j$ are connected by an edge and 0 otherwise.  On each node of the network there is a variable or spin~$s_i$, which is restricted to some discrete set of values~$S$.  In a compartmental model of disease propagation, for instance, $s_i\in S= \{0\;(\text{susceptible}),1\;(\text{infected}),\;2\;(\text{removed})\}$ could be the infection state of a node~\citei{kermack1927contribution,durrett1995spatial}.  In a spatial model of segregation $s_i\in S=\{0\;(\text{unoccupied}),\;1\;(\text{occupied})\}$ could represent land occupation~\citei{stauffer2007ising}.

Spins $s_i$ and $s_j$ interact if and only if there is an edge between nodes~$i$ and~$j$, a formulation sufficiently general to describe a large number of models in fields as diverse as statistical physics, machine learning, economics, psychology, epidemiology, and sociology \citei{pelizzola2005cluster,geman1986markov,yasuda2006triangular,decelle2011asymptotic,zhou2007self,galam1997rational,stauffer2008social}.  Interactions are represented by an interaction energy $g_{ij}(s_i,s_j| \omega_{ij})$, which controls the preference for any particular pair of states $s_i$ and $s_j$ to occur together.  The quantity $\omega_{ij}$ represents any external parameters, such as temperature in a classical spin system or infection rate in an epidemiological model, that control the nature of the interaction.  We also allow for the inclusion of an external field $f_i(s_i|\theta_i)$ with parameters~$\theta_i$, which controls the intrinsic propensity for~$s_i$ to take an particular state.  This could be used for instance to encode individual risk of catching a disease in an epidemic model.

Given these definitions, we write the probability~$P(\vec{s}|\omega,\theta)$ that the complete set of spins takes value~$\vec{s}$ in the Boltzmann form
\begin{align}
\label{state_prob}
P(\vec{s}|\omega,\theta) = \frac{e^{-H(\vec{s}| \omega,\theta)}}{Z(\omega,\theta)},
\end{align}
where the Hamiltonian
\begin{align}
\label{H_orig}
H(\vec{s}| \omega,\theta) = -\sum_{(i,j)\in E}g_{ij}(s_i,s_j| \omega_{ij})-\sum_{i\in V}f_i(s_i|\theta_i)
\end{align}
is the log-probability of the state to within an arbitrary additive constant, and the partition function
\begin{align}
\label{partition}
Z(\omega,\theta) = \sum_{\vec{s}}e^{-H(\vec{s}| \omega,\theta)}
\end{align}
is the appropriate normalizing constant, ensuring that $P(\vec{s}|\omega,\theta)$ sums to unity.  In this paper we will primarily be concerned with computing the single-site (or one-point) marginal probabilities
\begin{align}
\label{marg_def}
q_i(s_i) = \sum_{\vec{s}\setminus s_i}P(\vec{s}| \omega,\theta),
\end{align}
where $\vec{s}\setminus s_i$ denotes all spins with the exception of~$s_i$.  For convenience we have dropped $\omega$ and $\theta$ from the notation on the left of the equation, but it should be clear contextually that $q_i$ depends on both of these variables.

The one-point marginals reveal useful information about physical systems, such as the magnetization of classical spin models or the position of a phase transition.  They are important for statistical inference problems, where they give the posterior probability of a variable taking a given state after averaging over contributions from all other variables (e.g.,~the total probability of an individual being infected with a disease at a given time).  Unfortunately, direct computation of one-point marginals is difficult because the number of terms in the sum in Eq.~\eqref{marg_def} grows exponentially with the number of spins.  The message passing method gives us a way to get around this difficulty and compute $q_i$ accurately and rapidly.

Message passing can also be used to calculate other quantities.  For instance, we will show how to compute the average energy (also called the internal energy), which is given by
\begin{align}
U(\omega,\theta) = \sum_{\vec{s}}H(\vec{s}| \omega,\theta)P(\vec{s}| \omega,\theta).
\end{align}
The average energy is primarily of interest in thermo\-dynamic calculations, although it may also be of interest for statistical inference, where it corresponds to the average log-likelihood.

We can also compute the two-point correlation function between spins
\begin{align}
\label{two_pt}
P(s_i=x,s_j=y) = P(s_j=y| s_i=x)\,q_i(s_i=x).
\end{align}
This function can be computed by first calculating the one-point marginal $q_i(s_i=x)$, then fixing $s_i=x$ and repeating the calculation for~$s_j$.  The same approach can also be used to compute $n$-point correlation functions.

\subsection{Message passing equations}
\begin{figure}
\centering
\includegraphics[width=\columnwidth]{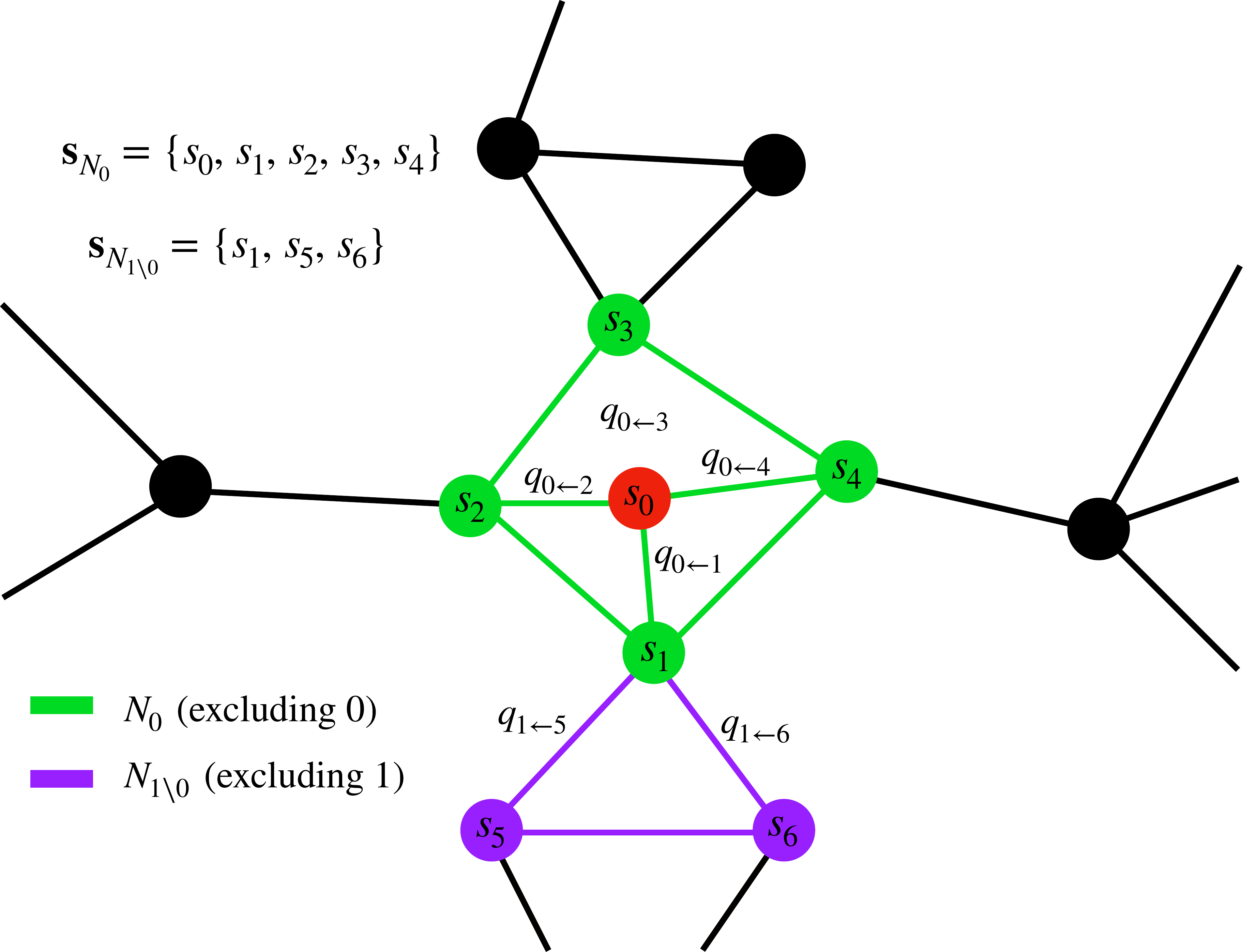}
\caption{\textbf{Hamiltonian expansion diagram, with $r=2$}.  The focal node is in red while the rest of its neighborhood~$N_0$ is in green.  Nodes and edges in purple represent the neighborhood $N_{1\setminus 0}$ excluding node~1.  We also label the corresponding spin and message variables used in Eqs.~\eqref{qi_MP} and~\eqref{qij_MP}.}
\label{neig_expansion}
\end{figure}

Our method operates by dividing a network into neighborhoods~\citei{cantwell2019message}.  A neighborhood $N_i^{(r)}$ around node~$i$ is defined as the node~$i$ itself and all of its edges and neighboring nodes, plus all nodes and edges along paths of length $r$ or less between the neighbors of~$i$.  See Fig.~\ref{neig_expansion} for examples.  The key to our approach is to focus initially on networks in which there are no paths longer than~$r$ between the neighbors of~$i$, meaning that all paths are inside~$N_i^{(r)}$.  This means that all correlations between spins within~$N_i^{(r)}$ are accounted for by edges that are also within~$N_i^{(r)}$, which allows us to write exact message passing equations for these networks.  Equivalently, we can define a \textit{primitive cycle} of length~$r$ starting at node~$i$ to be a cycle (i.e.,~a self-avoiding loop) such that at least one edge in the cycle is not on any shorter cycle beginning and ending at~$i$.  Our methods are then exact on any network that contains no primitive cycles of length greater than~$r+2$.

This approach gives us a series of methods where the $r$th member of the series is exact on networks that contain primitive cycles of length $r+2$ and less only.  The calculations become progressively more complex as $r$ gets larger: they are very tractable for smaller values but become impractical when $r$ is large.  In many real-world networks the longest primitive loop will be relatively long, requiring an infeasible computation to reach an exact solution.  However, long loops introduce smaller correlations between variables than short ones, and moreover the density of long loops is in many cases low: the network is ``locally dense but globally sparse.''  In this situation, we find that the message passing equations for low values of~$r$, while not exact, give excellent results.  They account correctly for the effect of the short loops in the network, while making only a small approximation by omitting the long ones.

In practice, quite modest values of $r$ can give good results.  The smallest possible choice is $r=0$, which corresponds to assuming that there are no loops in the network at all, that the network is a tree.  This is the assumption made by traditional message passing methods, and it gives poor results on many real-world networks.  The next approximation after this, however, with $r=1$, which correctly accounts for the effect of loops of length three in the network (i.e.,~triangles), produces substantially better results, and the $r=2$ approximation (which includes loops of length three and four) is in many cases impressively accurate.  In the following developments, we drop $r$ from our notation for convenience---the same equations apply for all values of~$r$.

Having defined the initial neighborhood~$N_i$ we further define a neighborhood~$N_{j\setminus i}$ to be node~$j$ plus all edges in $N_j$ that are not contained in~$N_i$ and the nodes at their ends.  Our method involves writing the marginal probability distribution on the spin at node~$i$ in terms of a set of messages received from nodes~$j$ that are in~$N_i$, including nodes that are not immediate neighbors of~$i$.  (This contrasts with traditional message passing in which messages are received only from the immediate neighbors of~$i$.)  These messages are then in turn calculated from further messages $j$ receives from nodes $k\in N_{j\setminus i}$, and so forth.

When written in this manner, the messages $i$ receives are independent of one another in any network with no primitive cycles longer than~$r+2$.  Messages received from any two nodes~$j_1$ and $j_2$ within $N_i$ are necessarily independent since they are calculated from the corresponding neighborhoods $N_{j_1\setminus i}$ and $N_{j_2\setminus i}$ which are disconnected from one another: if they were connected by any path then that path would create a primitive cycle starting at $i$ but passing outside of $N_i$, of which by hypothesis there are none.  By the same argument, we also know that $N_{j\setminus i}$ and $N_i$ only overlap at the single node~$j$ for any $j\in N_i$.

This much is in common with our previous approach in Ref.~\citei{cantwell2019message}, but to apply these ideas to the solution of probabilistic models we need to go further.  Specifically, we now show how this neighborhood decomposition allows us to factorize the Hamiltonian into a product of independent sums over the individual neighborhoods, with interactions that can be represented by messages passed between neighborhoods.  Consider $N_i$ as comprising a central set of nodes and edges surrounding~$i$.  Then we can think of the set of neighborhoods $N_{j\setminus i}$ for all $j\in N_i$ as comprising the next ``layer'' in the network, the sets $N_{k\setminus j}$ for all $k\in N_{j\setminus i}$ as a third layer, and so forth until all nodes and edges in the network are accounted for.  In a network with no primitive cycles longer than~$r+2$, this procedure counts all interactions exactly once, allowing us to rewrite our Hamiltonian as a sum of independent contributions from the various layers thus:
\begin{align}
\begin{split}
H(\vec{s}) &= H_{N_i}(\vec{s}_{N_i})
   + \sum_{j\in N_i} H_{N_{j\setminus i}}(\vec{s}_{N_{j\setminus i}}) \\
&\qquad{} + \sum_{j\in N_i} \, \sum_{k\in N_{j\setminus i}} H_{N_{k\setminus j}}(\vec{s}_{N_{k\setminus j}}) \\
&\qquad{} + \sum_{j\in N_i} \, \sum_{k\in N_{j\setminus i}} \, \sum_{l\in N_{k\setminus j}} H_{N_{l\setminus k}}(\vec{s}_{N_{l\setminus k}})
   + \ldots,
\end{split}
\label{factored_H}
\end{align}
where $\vec{s}_{N_i}$ and $\vec{s}_{N_{j\setminus i}}$ are the sets of spins for the nodes in the neighborhoods $N_i$ and~$N_{j\setminus i}$ and we have defined the local Hamiltonians
\begin{align}
\label{H_i}
H_{N_{i}}(\vec{s}_{N_{i}}) &= -\!\sum_{(j,k)\in N_{i}}g_{jk}(s_j,s_k| \omega_{jk})-f_i(s_i|\theta_i), \\
\label{H_cav}
H_{N_{j\setminus i}}(\vec{s}_{N_{j\setminus i}}) &= -\!\!\!\sum_{(k,l)\in N_{j\setminus i}}g_{kl}(s_k,s_l| \omega_{kl})-f_j(s_j|\theta_j).
\end{align}
The decomposition of Eq.~\eqref{factored_H} is illustrated pictorially in Fig.~\ref{neig_expansion}.

The essential feature of this decomposition is that it breaks sums over spins such as those in Eqs.~\eqref{partition} and~\eqref{marg_def} into a product of sums over the individual neighborhoods $\{N_{j\setminus i}\}_{j\in N_i}$.  Because these neighborhoods are, as we have said, independent, this means that the partition function and related quantities factorize into products of sums over a few spins each, which can easily be performed numerically.  For instance, the one-point marginal of Eq.~\eqref{marg_def} takes the form
\begin{align}
q_i(s_i=x) &\propto \!\!\!\sum_{\vec{s}_{N_i}:s_i=x} \!\!\! e^{- H_{N_i}(\vec{s}_{N_i})}\prod_{j\in N_i} \, \sum_{\vec{s}_{N_{j\setminus i}\setminus j}} \!e^{- H_{N_{j\setminus i}}(\vec{s}_{N_{j\setminus i}})} \nonumber\\
&\quad{} \times \prod_{k\in N_{j\setminus i}} \, \sum_{\vec{s}_{N_{k\setminus j}\setminus k}} \!\!e^{- H_{N_{k\setminus j}}(\vec{s}_{N_{k\setminus j}})}\ldots,
\end{align}
which can be written recursively as
\begin{align}
\label{qi_MP}
q_i(s_i=x) = \frac{1}{Z_i}\sum_{\vec{s}_{N_i}:s_i=x}e^{- H_{N_i}(\vec{s}_{N_i})}\prod_{j\in N_i}q_{i\leftarrow j}(s_j),
\end{align}
with
\begin{align}
\begin{split}
\label{qij_MP}
q_{i\leftarrow j}(s_j=y) &= \frac{1}{Z_{i\leftarrow j}}\sum_{\vec{s}_{N_{j\setminus i}}:s_j=y}e^{-H_{N_{j\setminus i}}(\vec{s}_{N_{j\setminus i}})} \\
&\hspace{6em}{} \times\prod_{k\in N_{j\setminus i}\setminus j}q_{j\leftarrow k}(s_k),
\end{split}
\end{align}
where the normalization constants $Z_i$ and $Z_{i\leftarrow j}$ ensure that the marginals $q_i$ and messages $q_{i\leftarrow j}$ are normalized so that they sum to $1$.  (In practice, we simply normalize the messages by dividing by their sum.)  The quantity~$q_{i\leftarrow j}(s_j)$ is equal to the marginal probability of node~$j$ having spin~$s_j$ when all the edges in~$N_i$ are removed.  Alternatively, one can think of it as a local external field on node~$j$ that influences the probability distribution of~$s_j$.  To make this more explicit one could rewrite Eq.~\eqref{qi_MP} as
\begin{align}
q_i(s_i=x) = \frac{1}{Z_i}\sum_{\vec{s}_{N_i}:s_i=x}e^{- H_{N_i}(\vec{s}_{N_i})+\sum_{j\in N_i}\log q_{i\leftarrow j}(s_j)},
\end{align}
where $\log q_{i\leftarrow j}(s_j)$ plays the role of the external field.

Equations~\eqref{qi_MP} and~\eqref{qij_MP} define our message passing algorithm and can be solved for the messages~$q_{i\leftarrow j}$ by simple iteration, starting from any suitable set of starting values and applying the equations repeatedly until convergence is reached.

With only slight modification we can use the same approach to calculate the internal energy as well.  The contribution to the internal energy from the interactions of a single node~$i$ is $\tfrac12\sum_{j:A_{ij}=1} g(s_i,s_j|\omega_{ij}) + f(s_i|\theta_i)$, where the factor of $\tfrac12$ compensates for double counting of interactions.  Summing over all nodes~$i$ and weighting by the appropriate Boltzmann probabilities, the total internal energy is
\begin{align}
U &= \sum_{i\in V} {1\over Z_i}
    \sum_{\vec{s}_{N_i}} \biggl[ \tfrac12\!
    \sum_{j:A_{ij}=1} g(s_i,s_j|\omega_{ij})
      + f(s_i|\theta_i) \biggr] \nonumber\\
  &\hspace{8em}{} \times e^{-H_{N_i}(\vec{s}_{N_i})}
            \prod_{j\in N_i} q_{i\leftarrow j}(s_j).
\label{U_MP}
\end{align}
All of the quantities appearing here are known a priori, except for the messages~$q_{i\leftarrow j}(s_j)$ and the normalizing constants~$Z_i$, which are calculated in the message passing process.  Performing the message passing and then using the final converged values in Eq.~\eqref{U_MP} then gives us our internal energy.

\subsection{Implementation}
\label{mcmc_desc}
For less dense networks, those with node degrees up to about~20, the message passing equations of Eqs.~\eqref{qi_MP} and~\eqref{qij_MP} can be implemented directly and work well.  The method is also easily parallelizable, as we can update all messages asynchronously based on their values from the previous iteration, as well as compute the final marginals in parallel.

For networks with higher degrees the calculations can become unwieldy, the huge reduction in complexity due to the factorization of the Hamiltonian notwithstanding.  For a model with $t$ distinct spin states at every node, the sum over states in the neighborhood of~$i$ has $t^{\abs{N_i}}$ terms, which can quickly become computationally expensive to evaluate.  Moreover, if just a single node has too large a neighborhood it can make the entire computation intractable, as that single neighborhood can consume more computational power than is available.

In such situations, therefore, we take a different approach.  We note that Eq.~\eqref{qij_MP} is effectively an expectation
\begin{align}
\label{avg_qs_cav}
q_{i\leftarrow j}(s_j=y) = \langle \delta_{s_j,y}\rangle_{N_{j\setminus i}},
\end{align}
where
we use the shorthand
\begin{align}
\label{shorthand}
\langle A\rangle_{N_{j\setminus i}} = \sum_{\vec{s}_{N_{j\setminus i}}}A(\vec{s}_{N_{j\setminus i}})\frac{e^{-H_{N_{j\setminus i}}(\vec{s}_{N_{j\setminus i}})}\prod_{k\in N_{j\setminus i}\setminus j}q^{j\leftarrow k}_{s_k}}{Z^{i\leftarrow j}}.    
\end{align}
We approximate this average using Markov chain Monte Carlo importance sampling over spin states, and after convergence of the messages the final estimates of the marginals $q_i$ can also be obtained by Monte Carlo, this time on the spins in~$N_i$.  We describe the details in Section~\ref{applications}.

\subsection{Calculating the partition function}
\label{main:neig_factorization}
The partition function $Z$ is perhaps the most fundamental quantity in equilibrium statistical mechanics.  From a knowledge of the partition function one can calculate virtually any other thermodynamic variable of interest.  Objects equivalent to $Z$ also appear in other fields, such as Bayesian statistics, where the quantity known as the \emph{model evidence}, the marginal likelihood of observed data given a hypothesized model, is mathematically analogous to the partition function and plays an important role in model fitting and selection~\citei{mackay2003information, migon2014statistical,friel2012estimating}.

Unfortunately, the partition function is difficult to calculate in practice.  The calculation can be done analytically in some special cases~\citei{salinas2001introduction,baxter2016exactly}, but direct numerical calculations are difficult due to the need to sum over an exponentially large number of states, and Monte Carlo is challenging because of the difficulty of correctly normalizing the Boltzmann distribution.

Another concept central to statistical mechanics is the entropy
\begin{align}
S = - \sum_{\vec{s}} P(\vec{s}) \ln P(\vec{s}),
\label{entropy}
\end{align}
which has broad applications not just in physics but across the sciences~\citei{shannon1951prediction, bialek2012biophysics, cabral2013entropy}.  Like the partition function, entropy is difficult to calculate numerically, and for exactly the same reasons, since the two are closely related.  For the canonical distribution of Eq.~\eqref{state_prob} the entropy is given in terms of $Z$ by $S = \ln Z + \beta U$.  Even if we know the internal energy~$U$ therefore (which is relatively straightforward to compute), the entropy is at least as hard to calculate as the partition function.  Indeed the fundamental difficulty of normalizing the Boltzmann distribution is equivalent to establishing the zero of the entropy, a well known hard problem (unsolvable within classical thermodynamics, requiring the additional axiom of the Third Law).

As we now show, the entropy can be calculated using our message passing formalism by appropriately factorizing the probability distribution over spin states. Since we have already developed a prescription for computing~$U$ (see Eq.~\eqref{U_MP}), this also allows us to calculate the partition function. The details of the procedure are quite involved and do not follow straightforwardly from the previous discussion, so we defer the derivation to the Supporting Materials, Section~IV.  As shown there, the state probability~$P(\vec{s})$ in Eq.~\eqref{state_prob} can be rewritten in the factorized form
\begin{align}
P(\vec{s}) = { \prod_{i \in G} P(\vec{s}_{{N}_i}) \over \prod_{((i,j)) \in G} P(\vec{s}_{\cap_{ij}})^{2 / \vert \cap_{ij} \vert}},
\label{neighborhood-ansatz-1}
\end{align}
where $P(\vec{s}_{N_i})$ is the joint marginal distribution of the variables in the neighborhood of node~$i$, $P(\vec{s}_{\cap_{ij}})$ is the joint marginal distribution in the intersection $\cap_{ij} = N_i \cap N_j$ of the neighborhoods~$N_i$ and~$N_j$, and $((i,j))$ denotes pairs of nodes that are contained in each other's neighborhoods.

By a series of manipulations, this form can be further expressed as the pure product 
\begin{align}
P(\vec{s}) &=  \biggl[ \prod_{((i,j)) \in G} P( \vec{s}_{\cap_{ij}})^{1 / {\vert \cap_{ij} \vert \choose 2}} \biggr] \biggl[ \prod_{(i,j) \in G} P(s_i,s_j)^{W_{ij}} \biggr] \nonumber\\
&\qquad\times \biggl[ \prod_{i \in G} P(s_i)^{C_i} \biggr],
\label{N_factorization}
\end{align}
where
\begin{align}
	W_{ij} = 1 - \sum_{ ((l,m)) \in G } { 1 \over { \vert \cap_{lm} \vert \choose 2 }} \mathbf{1}_{ \{ (i,j) \in \cap_{lm} \} }
\end{align}
with $\mathbf{1}_{ \{ \dots \} }$ being the indicator function, and
\begin{align}
	C_i = 1 - \Bigg( \sum_{j \in {N}_i} {1 \over \vert \cap_{ij} \vert -1} \Bigg) - \Bigg( \sum_{j \in N_i^{(0)}} W_{ij}  \Bigg).
\end{align}
Substituting Eq.~\eqref{N_factorization} into Eq.~\eqref{entropy}, we get an expression for the entropy thus:
\begin{align}
&S = -{1 \over {\vert \cap_{ij} \vert \choose 2}  }\sum_{((i,j)) \in G} \!\!\!
P( \vec{s}_{\cap_{ij}})\ln P( \vec{s}_{\cap_{ij}}) \nonumber\\ 
&-\!\!\sum_{(i,j) \in G}\!\!\! W_{ij} P(s_i, s_j )\ln P(s_i, s_j )
- \sum_{i \in G} C_i  P( s_i)\ln P( s_i).
\label{neighborhood-entropy}
\end{align}

Note that, like the well known Bethe approximation for the entropy~\citei{yedidia2003understanding}, this expression has contributions from the one- and two-point marginals $P(s_i)$ and $P(s_i,s_j)$ of Eqs.~\eqref{two_pt} and~\eqref{qi_MP}, but also contains a term that depends on the joint marginal~$P( \vec{s}_{\cap_{ij}})$ in the intersection $\cap_{ij}$, which may be nontrivial if $r>0$.  As shown in the Supporting Materials, Section IV, we can calculate this joint marginal using the message passing equation
\begin{equation}
P(\vec{s}_{\cap_{ij}}) = \frac{1}{Z_{\cap_{ij}}} e^{-\beta H(\vec{s}_{\cap_{ij}})} q_{i 
\leftarrow j}(s_j) \prod_{k \in \cap_{ij} \setminus j } q_{j \leftarrow k} (s_k),
\end{equation}
where $H(\vec{s}_{\cap_{ij}})$ denotes the terms of the Hamiltonian of Eq.~\eqref{H_orig} that fall within~$\cap_{ij}$ and $Z_{\cap_{ij}}$ is the corresponding normalizing constant.  For $|\cap_{ij}|$ sufficiently small, $Z_{\cap_{ij}}$~can be computed exactly.  In other cases we can calculate it using Monte Carlo methods similar to those we used previously for the marginals~$P(s_i)$.

\subsection{Ising model calculations}
As an archetypal application of our methods we consider the Ising model on various example networks.  The ferromagnetic Ising model in zero external field is equivalent in our notation to the choices
\begin{equation}
g_{ij}(s_i,s_j) = -\beta A_{ij} s_is_j,\qquad f_i(s_i) = 0,
\label{isingfg}
\end{equation}
where $\beta=1/T$ is the inverse temperature.  Note that temperature in this notation is considered a part of the Hamiltonian.  It is more conventional to write temperature separately, so that the Hamiltonian has dimensions of energy rather than being dimensionless as here, but absorbing the temperature into the Hamiltonian is notationally convenient in the present case.  It effectively makes the temperature a parameter~$\omega_{ij}$ in Eq.~\eqref{H_orig} (and all $\omega_{ij}$ are equal).

As example calculations, we will compute the average magnetization~$M$, which is given by
\begin{align}
M = \left|\biggl\langle\frac{1}{N}\sum_{i=1}^{N}s_i \biggr\rangle \right|
  = \frac{1}{N}\left|\sum_{i=1}^N \bigl[2q_i(s_i=+1)-1 \bigr] \right|,
\label{magnetization}
\end{align}
and the heat capacity~$C$, given by
\begin{align}
C = \frac{dU}{dT}=-\beta^2\frac{dU}{d\beta}.
\label{C_identity}
\end{align}
As detailed in Section I of the Supporting Materials, we employ an extension of the message passing equations to compute~$C$ that avoids having to use a numerical derivative to evaluate Eq.~\eqref{C_identity}.  In brief, we consider the messages $q_{i\leftarrow j}$ to be a function of $\beta$ then define their derivatives with respect to $\beta$ as their own set of messages
\begin{align}
\eta_{i\leftarrow j} = \frac{dq_{i\leftarrow j}}{d\beta},    
\end{align}
with their own associated message passing equations derived by differentiating Eq.~\eqref{qij_MP}.  We then compute the heat capacity~$C$ by differentiating Eq.~\eqref{U_MP}, expressing the result in terms of the~$\eta_{i\leftarrow j}$, and substituting it into Eq.~\eqref{C_identity}.

\subsection{Behavior at the phase transition}
In many geometries, the ferromagnetic Ising model has a phase transition at a nonzero critical temperature between a symmetric state with zero average magnetization and a symmetry broken state with nonzero magnetization.  Substituting Eq.~\eqref{isingfg} into Eqs.~\eqref{qi_MP} and~\eqref{qij_MP} we can show that the message passing equations for the Ising model always have a trivial solution $q_{i\leftarrow j}(s_j) = \frac12$ for all~$i,j$.  This choice is a fixed point of the message passing iteration: when started at this point the iteration will remain there indefinitely.  Looking at Eq.~\eqref{magnetization}, we see that this fixed point corresponds to magnetization $M=0$.  If the message passing iteration converges to this trivial fixed point, therefore, it tells us that the magnetization is zero and we are above the critical temperature; if it settles elsewhere then the magnetization is nonzero and we are below the critical temperature.  Thus the phase transition corresponds to the point at which the fixed point changes from being attracting to being repelling.

This behavior is well known in standard belief propagation, where it has been shown that on networks with long loops only there is a critical temperature~$T_\textrm{BP}$ below which the trivial fixed point becomes unstable and hence the system develops nonzero magnetization, and that this temperature corresponds precisely to the conventional zero-field continuous phase transition on these networks~\citei{mooij2005properties}.  Extending the same idea to the present case, we expect the phase transition on a loopy network to fall at the corresponding transition point between stable and unstable in our message passing formulation.

Moreover, because the values of the messages at the trivial fixed point are known, we can compute an expression for the phase transition point without performing any message passing.  We treat the message passing iteration as a dynamical system and perform a linear stability analysis of the trivial fixed point.  Perturbing around $q=\frac{1}{2}$ (shorthand for setting all $q_{i\leftarrow j} = \frac12$) and keeping terms to linear order, we find that the dynamics is governed by the Jacobian
\begin{align}
J_{j\to i,\nu\to\mu}= \frac{\partial q_{i\leftarrow j}}{\partial q_{\mu\leftarrow \nu}}\Big\vert_{q=1/2} = \tilde B_{j\to i,\nu\to\mu}D_{j\to i,\nu\to\mu},
\label{jacobian}
\end{align}
where $\tilde B$ is a generalization of the so-called non-backtracking matrix~\citei{Krzakala13} to our loopy message passing formulation:
\begin{align}
\tilde B_{j\to i,\nu\to\mu} = \biggl\lbrace\begin{array}{ll}
  1 & \quad\mbox{if $j=\mu$ and $\nu\in N_{j\setminus i}$,}\\
  0 & \quad\mbox{otherwise,}
\end{array}
\label{eq:nonbacktracking}
\end{align}
and $D_{j\to i,\nu\to\mu}$ is a correlation function between the spins $s_\mu$ and $s_\nu$ within the neighborhood $N_{j\setminus i}$---see Section III of the Supporting Materials for details.

When the magnitude of the leading eigenvalue $\lambda_\textrm{max}$ of this Jacobian is less than~1, the trivial fixed point is stable; when it is greater than~1 the fixed point is unstable.  Hence we can locate the phase transition temperature by numerically evaluating the Jacobian and locating the point at which $\abs{\lambda_\textrm{max}}$ crosses~1, for instance by binary search.

Equation~\ref{eq:nonbacktracking} is also useful in its own right.  The non-backtracking matrix has numerous applications within network science, for instance in community detection~\citei{Krzakala13}, centrality measures~\citei{MZN14}, and percolation theory~\citei{KNZ14}.  The generalization defined in Eq.~\ref{eq:nonbacktracking} could be used to extend these applications to loopy networks, although we will not explore such calculations here.

\section{Results}
\label{applications}
\subsection{A model network}
\label{sec:artificial}
As a first example application, we examine the behavior of our method on a model network created precisely to have short loops only up to a specified maximum length.  The network has short primitive cycles only of length $r+2$ and less for a given choice of~$r$, though it can also have long loops---it is ``locally dense but globally sparse'' in the sense discussed previously.  Indeed this turns out to be a crucial point.  The Ising model does not have a normal phase transition on a true tree, because at any finite temperature there is always a nonzero density of defects in the spin state (pairs of adjacent spins that are oppositely oriented), which on a tree divide the network into finite sized regions, imposing a finite correlation length and hence no critical behavior.  Similarly in the case of a network with only short loops and no long ones there is no true phase transition.  The long loops are necessary to produce criticality, a point discussed in detail in~\citei{allard2019accuracy}.

To generate networks that have short primitive cycles only up to a certain length, we first generate a random bipartite network---a network with two types of nodes and connections only between unlike kinds---then ``project'' down onto one type of node, producing a network composed of a set of complete subgraphs or cliques.  In detail, the procedure is as follows.
\begin{enumerate}
  \item We first specify the degrees of all the nodes, of both types, in the bipartite network.
  \item We represent these degrees by ``stubs'' of edges emerging, in the appropriate numbers, from each node, then we match stubs at random in pairs to create our random bipartite network.
  \item We project this network onto the nodes of type~1, meaning that any two such nodes that are both connected to the same neighbor of type~2 are connected directly with an edge in the projection.  After all such edges have been added, the type-2 nodes are discarded.
  \item Finally, we remove a fraction $p$ of the edges in the projected network at random.
\end{enumerate}

If $p=0$, the network is composed of fully connected cliques, but when $p>0$ some cliques will be lacking some edges, and hence the network is composed of a collection of subgraphs of size equal to the degrees of the corresponding nodes of type~2 from which they were projected.  If we limit these degrees to a maximum value of~$r+2$ then there will be no short loops of length longer than this.

\begin{figure}
  \centering
  \includegraphics[width=.49\textwidth]{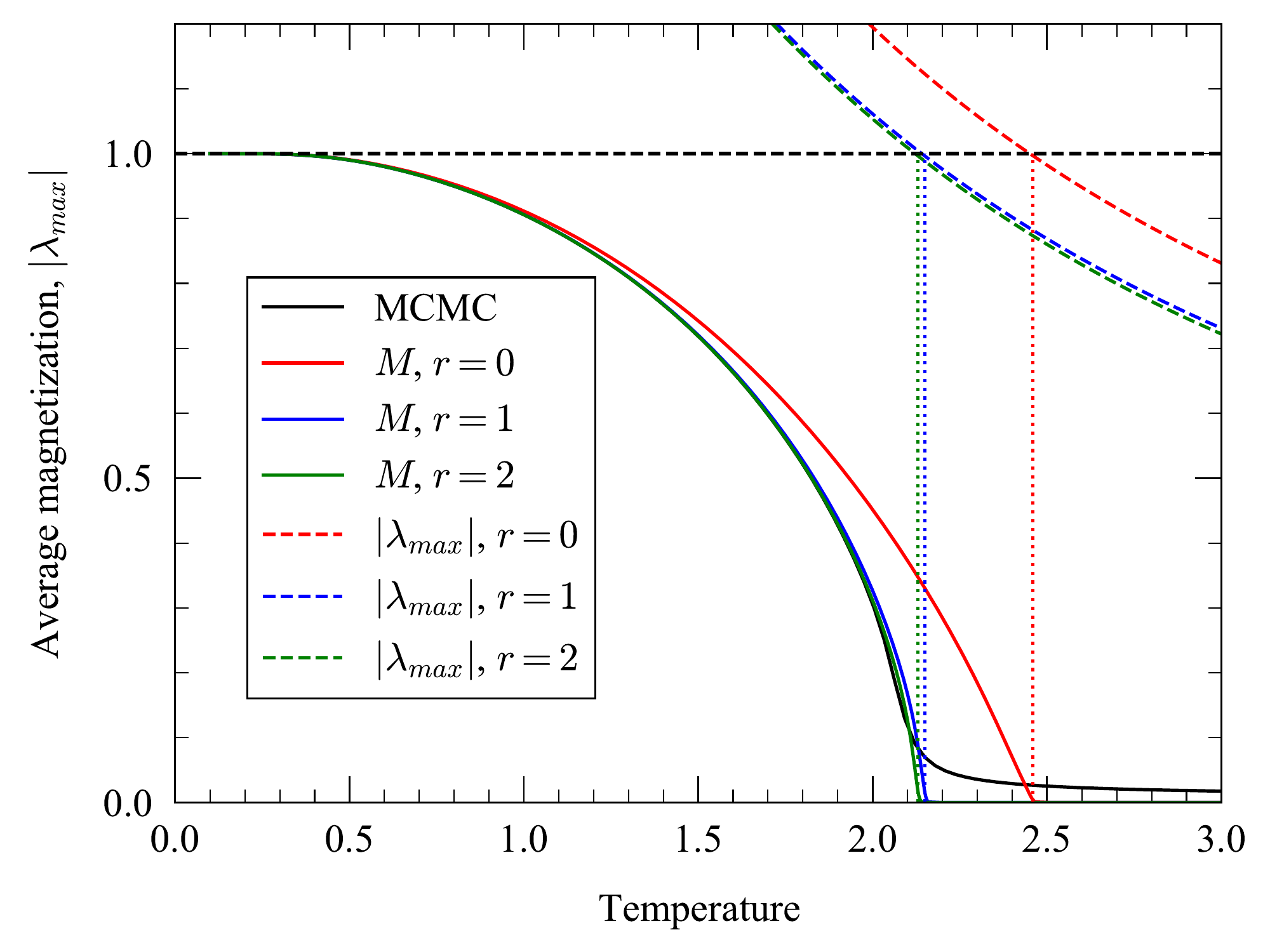}
  \includegraphics[width=.49\textwidth]{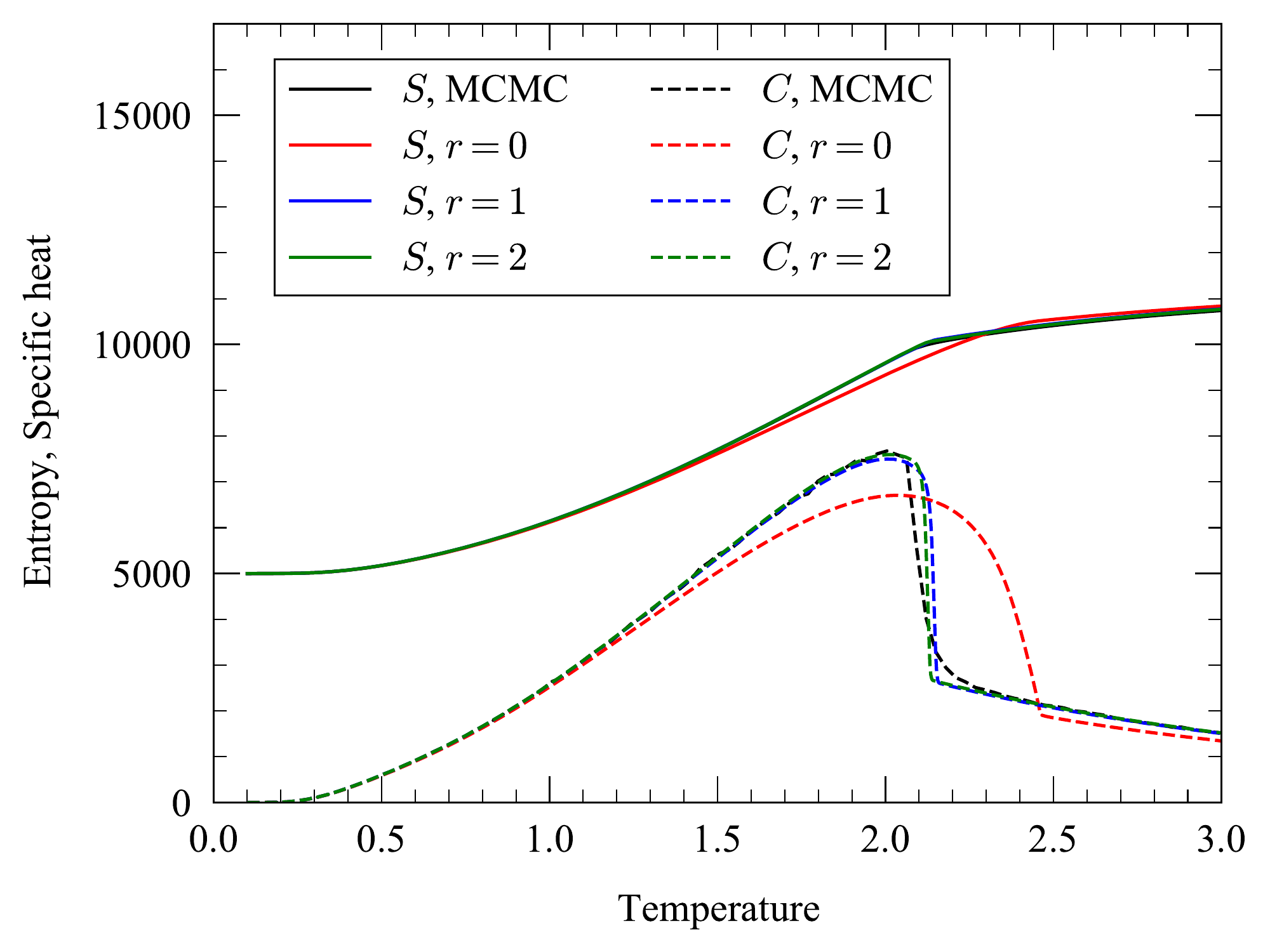}
  \caption{\textbf{Ferromagnetic Ising model critical behavior on synthetic network.} The top panel shows the average magnetization, while the bottom one shows the heat capacity and the entropy (the latter shifted up for visualization purposes).  The magnitude of the leading eigenvalue for the Jacobian is also shown in the top panel for all three values of~$r$, and we can see that the apparent positions of the phase transition, revealed by discontinuities in the physical quantities or their gradients, correspond closely to the temperatures at which the associated eigenvalues are equal to~1.}
  \label{large_exact}
\end{figure}

Figure~\ref{large_exact} shows the magnetization per spin, entropy, and heat capacity for the ferromagnetic Ising model on an example network of $9\,447$ nodes and $13\,508$ edges generated using this procedure with $r=2$ and $p = 0.6$.  We also limit the degrees of the type-1 nodes in the bipartite graph to a maximum of~5, which ensures that no neighborhood in the projection is too large to prevent a complete summation over states and hence that Monte Carlo estimation of the sums in the message passing equations is unnecessary.

Results are shown for belief propagation calculations with $r=0$, 1, and~2, the last of which should, in principle, be exact except for the weak correlations introduced by the presence of long loops in the network.  We also show in the figure the magnitude of the leading eigenvalue of $J$ for each value of~$r$.  The points at which this eigenvalue equals~1, which give estimates of the critical temperature for each~$r$, are indicated by the vertical lines.  Also shown in the figure for comparison are results from direct Monte Carlo simulations of the system, with the entropy calculated from values of the heat capacity computed from energy fluctuations and then numerically integrated using the identity
\begin{align}
S = \int_0^T \frac{C(T)}{T}\>dT.
\end{align}
The message passing simulations offer significantly faster results for this system: for $r=2$ message passing was about 100 times faster than the Monte Carlo simulations.

Looking at Fig.~\ref{large_exact}, we can see that as we increase~$r$ the message passing results approach those from the direct Monte Carlo, except close to the phase transition, where the Monte Carlo calculations suffer from finite size effects that smear the phase transition, to which the message passing approach appears largely immune.  While the results for conventional belief propagation ($r=0$) are quite far from the direct Monte Carlo results, most of the improvement in accuracy from our method is already present even at $r=1$.  Going to $r=2$ offers only a small additional improvement in this case.

The apparent position of the phase transition aligns well with the predictions derived from the value of the Jacobian for each value of~$r$.  The transition is particularly clear in the gradient discontinuity of the magnetization.  For $r=1$ and~2 the heat capacity appears to exhibit a discontinuity at the transition, which differs from the divergence we expect on low-dimensional lattices but bears a resemblance to the behavior seen on Bethe lattices and other homogeneous tree-like networks~\citei{bethe1935statistical,mancini2006equations,dorogovtsev2008critical}.

\subsection{Real-world networks}
For our next example we look at an application on a real-world network, where we do not expect the method to be exact, though as we will see it nonetheless performs well.  The network we examine has larger local neighborhoods than our synthetic example, which means we are not able to sum exhaustively over all configurations of the spins~$\vec{s}_{N_{j\setminus i}}$ in Eq.~\eqref{qij_MP} (and similarly $\vec{s}_{N_i}$ in Eq.~\eqref{qi_MP}) so, as described in Section~\ref{mcmc_desc}, we instead make use of Monte Carlo sampling to estimate the messages~$q_{i\leftarrow j}$ and marginals~$q_i$.

The summation over local spins in Eq.~\eqref{qij_MP} is equivalent to computing the expectation in Eq.~\eqref{avg_qs_cav}.  To calculate $q_{i\leftarrow j}(s_j=y)$ we fix the values of its incoming messages $\{q_{j\leftarrow k}\}$ and perform Monte Carlo sampling over the states of the spins in the neighborhood $N_{j\setminus i}$ with the Hamiltonian of Eq.~\eqref{H_cav}.  Then we compute the average in Eq.~\eqref{avg_qs_cav} separately for the cases $y=1$ and $-1$ and normalize to ensure that the results sum to one.  The resulting values for $q_{i\leftarrow j}$ can then be used as incoming messages for calculating other messages in other neighborhoods.  We perform the Monte Carlo using the Wolff cluster algorithm~\citei{wolff1989collective}, which makes use of the Fortuin-Kasteleyn percolation representation of the Ising model to flip large clusters of spins simultaneously and can significantly reduce the time needed to obtain independent samples, particularly close to the critical point.  Once the messages have converged to their final values we compute the marginals~$q_i$ by performing a second Monte Carlo, this time over the spins~$\vec{s}_{N_i}$ with the Hamiltonian of Eq.~\eqref{H_i}.  More details on the procedure are given in Section II of the Supporting Materials.

The Monte Carlo approach combines the best aspects of message passing and traditional Monte Carlo calculations.  Message passing reduces the sums we need to perform to sets of spins much smaller than the entire network, while the Monte Carlo approach dramatically reduces the number of spin \emph{states} that need to be evaluated.  The approach has other advantages too.  For instance, because of the small neighborhood sizes it shows improved performance in systems with substantial energy barriers that might otherwise impede ergodicity, such as antiferromagnetic systems.  But perhaps its biggest advantage is that it effectively allows us to sample very large numbers of states of the network without taking very large samples of individual neighborhoods.  If we sample $k$ configurations from one neighborhood and $k$ configurations from another, then in effect we are summing over $k^2$ possible combinations of states in the union of the two neighborhoods.  Depending on the value of~$r$, there are at least $2m$ neighborhoods $N_{j\setminus i}$ in a network, where $m$ is the number of edges, and hence we are effectively summing over at least~$k^{2m}$ states overall, a number that increases exponentially with network size.  Effective sample sizes of $10^{1000}$ or more are easily reachable, far beyond what is possible with traditional Monte Carlo methods.

\begin{figure}
\centering
\includegraphics[width=.49\textwidth]{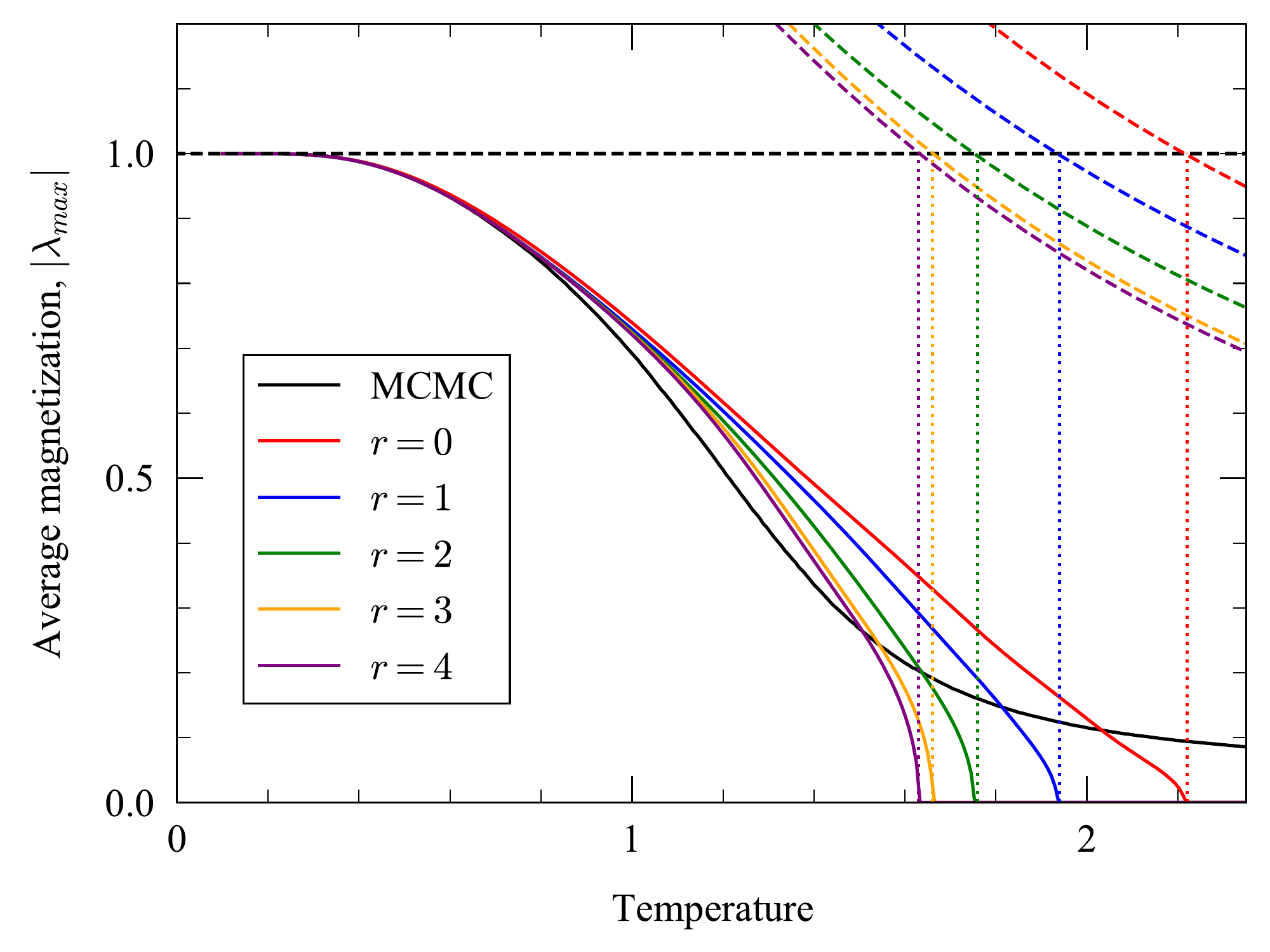}
\includegraphics[width=.49\textwidth]{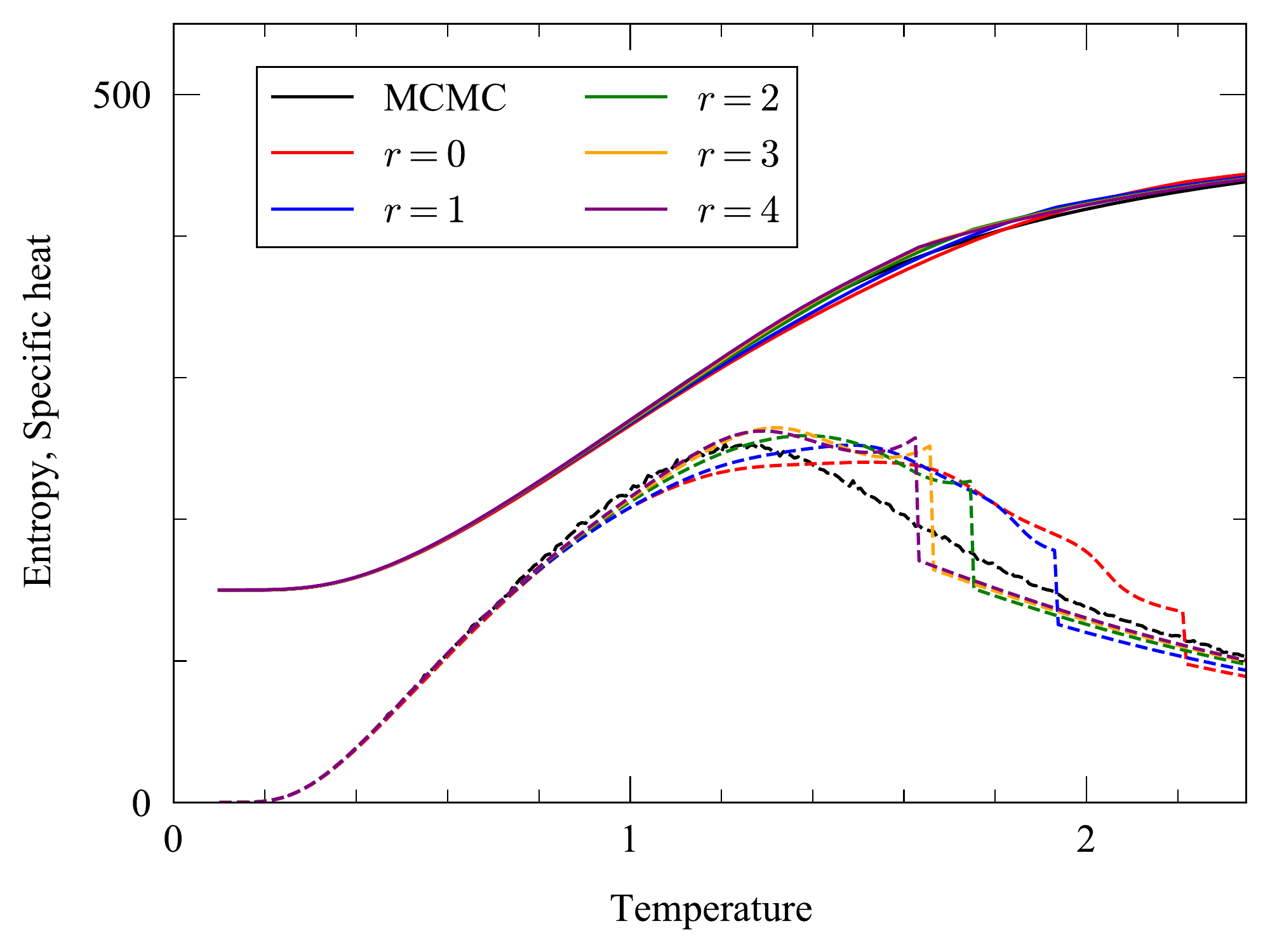}
\caption{\textbf{Ferromagnetic Ising model critical behavior on a power grid network.}  Message passing and Monte Carlo calculations of the average magnetization, entropy, and specific heat on the ``494 bus power system'' network from Ref.~\citei{davis2011university}.  Again, the message passing results approximate the real solution progressively better as~$r$ grows larger.}
\label{mcmc_fig}
\end{figure}

Figure~\ref{mcmc_fig} shows the results of applying these methods with $r=0\ldots4$ to a network from~\citei{davis2011university} representing the structure of an electric power grid, along with results from direct Monte Carlo simulations on the same network.  As the figure shows, the magnetization is again poorly approximated by the traditional ($r=0$) message passing algorithm, but improves as~$r$ increases.  In particular, the behavior in the region of the phase transition is quite poor for $r=0$ and does not provide a good estimate of the position of the transition.  For $r=1$ and~2, however, we get much better estimates, and for $r=3$ and~4 the method approaches the Monte Carlo results quite closely, with the critical temperature falling somewhere in the region of $T=1.6$ in this case.  We also see a much clearer phase transition in the message passing results than in the standard Monte Carlo, because of finite size effects in the latter.  These results all suggest that for real systems our method can give substantial improvements over both ordinary belief propagation and direct Monte Carlo simulation, and in some cases show completely different behavior altogether.

\section{Discussion}
In this paper we have presented a new class of message passing algorithms for solving probabilistic models on networks that contain a high density of short loops.  Taking the Ising model as an example, we have shown that our methods give substantially improved results in calculations of magnetization, heat capacity, entropy, marginal spin probabilities, and other quantities over standard message passing methods that do not account for the presence of loops.  Our methods are exact on networks with short loops up to a fixed maximum length which we can choose, and can give good approximations on networks with loops of any length.

Message passing methods for probabilistic models on loopy networks have been proposed in the past, the best known being the generalized belief propagation method of Yedidia~\etal~\citei{yedidia2001generalized}.  Generalized belief propagation employs a \emph{region-based approximation}~\citei{yedidia2005constructing}, in which the free energy $\ln Z$ is approximated by a sum of independent local free energies of regions within the network.  Once the regions are defined it is straightforward to write down belief propagation equations, which can be used to calculate marginals and other quantities of interest, including approximations to the partition function and entropy.  Perhaps the best known example of generalized belief propagation, at least within the statistical physics community, is the \emph{cluster variational method}, in which the regions are defined so as to be closed under the intersection operation~\citei{pelizzola2005cluster} and the resulting free energy is called the \emph{Kikuchi free energy}~\citei{kikuchi1951theory}.

The accuracy and complexity of generalized belief propagation is determined by the specific choice of regions, which has been described as being ``more of an art than a science''~\citei{yedidia2003understanding}.  Loops contained within regions are correctly accounted for in the belief propagation, while those that span two or more regions are not and introduce error.  At the same time, the computational complexity of the belief propagation calculations increases exponentially with the size of the regions~\citei{yedidia2003understanding}, so choosing the right regions is a balancing act between enclosing as many loops as possible while not making the regions too large.  A number of heuristics have been proposed for choosing the regions~\citei{kappen2002novel,pakzad2002minimal,welling2004choice} but real-world networks can pose substantial difficulties because they often contain both high degrees and many loops~\citei{newman2018networks}, which effectively forces us to compromise either by leaving loops out or by using very large regions.  Our method can have a significant advantage in these systems because it can accommodate large, tightly connected neighborhoods through local Monte Carlo sampling.  Our method also has the benefit that the neighborhoods are constructed automatically based on the network structure rather than being chosen by the user.

There are many ways in which the methods and results of this paper could be extended.  We have studied only one application in detail, the Ising model, but the formalism we present is a general one that could be applied to many other models.  In principle, any model with sparse pairwise interactions (i.e.,~interactions whose number scales sub-quadratically with the number of variables) could be studied using these methods.  For example, there is a large class of generative models of networks in which edges appear with probabilities that depend on the properties of the adjacent nodes.  Examples include the Chung-Lu model~\citei{CL02b} and the stochastic block model and its variants~\citei{HLL83,KN11a}.  If we assume an observed network to be drawn from such a model then we can use statistical inference to estimate the values of hidden node attributes that influence edge probability, such as community membership.  Our message passing methods could be applied to such inference calculations and could in principle give more accurate results in the common case where the observed network contains many short loops.

Another potential application in the realm of statistical inference is the inverse Ising model, the problem of inferring the parameters of an Ising or Ising-like model from an observed sequence of spin states, which has numerous applications including the reconstruction of neural pathways~\citei{schneidman2006weak}, the inference of protein structure~\citei{morcos2011direct}, and correlations within financial markets~\citei{bury2013market}.  It can be shown that the one- and two-point correlation functions of the observed spins are sufficient statistics to reliably estimate coupling and external field parameters~\citei{nguyen2017inverse} and our method could be used to compute these statistics on loopy networks to greater accuracy than with traditional message passing and faster than standard Monte Carlo simulation.  Other potential applications, further afield from traditional statistical physics, include the solution of constraint satisfaction problems, coding theory, and combinatorial optimization.

\subsection*{Acknowledgments}
This work was funded in part by the US Department of Defense NDSEG fellowship program (AK) and by the National Science Foundation under grants DMS--1710848 and DMS--2005899 (MEJN).\\




\bibliographystyle{numeric}

\clearpage
\appendix 

\section{Supporting Materials}

\subsection{Calculation of the heat capacity\\using message passing}
\label{SI:C_details}
The heat capacity, which is given by
\begin{equation}
\label{C_def}
C = \frac{dU}{dT} = -\beta^2 \frac{dU}{d\beta},
\end{equation}
can be calculated from the expression for the internal energy
\begin{align}
U(\beta) = \frac12 \sum\limits_{i\in V}\frac{1}{Z^i(\beta)}\sum\limits_{\vec{s}_{N_i}} H_{\partial_i}(\vec{s}_{\partial_i}) \,e^{-\beta H_{N_i}(\vec{s}_{N_i})}\nonumber \\
\times \prod\limits_{j\in N_i\setminus i} q^{i\leftarrow j}_{s_j}(\beta),
\label{app:U_MP}
\end{align}
where instead of incorporating the $\beta$ dependence into the Hamiltonian as in the main paper, we now display it explicitly.  In this expression, $N_i$~denotes the neighborhood of node~$i$ as in the main text, $\partial_i$~denotes the node~$i$ and its immediately adjacent edges and nodes, and $H_{N_{i}}(\vec{s}_{N_{i}})$ and $H_{\partial_i}(\vec{s}_{\partial_i})$ represent the terms in the Hamiltonian for these subgraphs:
\begin{align}
H_{N_{i}}(\vec{s}_{N_{i}}) = - f_i(s_i\vert\theta_i)
 - \!\!\!\sum\limits_{(j,k)\in N_{i}}g_{jk}(s_j,s_k\vert \omega_{jk})
\end{align}
and
\begin{align}
H_{\partial_i}(\vec{s}_{\partial_i}) = - 2f_i(s_i\vert\theta_i)
- \!\!\sum\limits_{(i,j)\in \partial_i}g_{ij}(s_i,s_j\vert \omega_{ij}),
\end{align}
with the $\beta$ dependence omitted from the definition of the functions.  With the $\beta$ dependence written in this way the message passing equations take the form
\begin{align}
\label{app:qi_MP}
q^i_x(\beta) = \frac{1}{Z^{i}(\beta)}\sum\limits_{\vec{s}_{N_i\setminus i}}\delta_{s_i,x}e^{-\beta H_{N_i}(\vec{s}_{N_i})}
\prod\limits_{j\in N_i\setminus i}q^{i\leftarrow j}_{s_j}(\beta),
\end{align}
and
\begin{align}
\label{app:qij_MP}
q^{i\leftarrow j}_{y}(\beta) = \frac{1}{Z^{i\leftarrow j}(\beta)}\sum\limits_{\vec{s}_{N_{j\setminus i}}}\delta_{s_j,y}e^{-\beta H_{N_{j\setminus i}}(\vec{s}_{N_{j\setminus i}})}\nonumber\\
\times\prod\limits_{k\in N_{j\setminus i}\setminus j}\!\! q^{j\leftarrow k}_{s_k}(\beta),
\end{align}
with
\begin{align}
\label{Zs}
Z^{i}(\beta) &= \sum\limits_{\vec{s}_{N_i}}e^{-\beta H_{N_i}(\vec{s}_{N_i})}\prod\limits_{j\in N_i\setminus i}q^{i\leftarrow j}_{s_j}(\beta), \\
Z^{i\leftarrow j}(\beta) &= \sum\limits_{\vec{s}_{N_{j\setminus i}}}e^{-\beta H_{N_{j\setminus i}}(\vec{s}_{N_{j\setminus i}})}
\!\!\prod\limits_{k\in N_{j\setminus i}\setminus j} \!\!q^{j\leftarrow k}_{s_k}(\beta).
\end{align}
Differentiating~\eqref{app:qij_MP} with respect to~$\beta$ and defining the quantity
\begin{align}
\eta_{y}^{i\leftarrow j} = \frac{dq^{i\leftarrow j}_{y}}{d\beta},
\end{align}
we get
\begin{align}
\label{dmessage}
\eta_{y}^{i\leftarrow j} &=  \frac{1}{Z^{i\leftarrow j}(\beta)}\sum\limits_{\vec{s}_{N_{j\setminus i}}}e^{-\beta H_{N_{j\setminus i}}(\vec{s}_{N_{j\setminus i}})}\prod\limits_{k\in N_{j\setminus i}\setminus j}q^{j\leftarrow k}_{s_k}(\beta)\nonumber\\
&\quad{} \times\biggl(\left[q^{i\leftarrow j}_{y}(\beta)-\delta_{s_j,y}\right]H_{N_{j\setminus i}}(\vec{s}_{N_{j\setminus i}})
\nonumber\\
&\quad{} + \left[\delta_{s_j,y}-q^{i\leftarrow j}_{y}(\beta)\right]\sum\limits_{k\in N_{j\setminus i}\setminus j}\frac{\eta^{j\leftarrow k}_{s_k}(\beta)}{q^{j\leftarrow k}_{s_k}(\beta)}\biggr),
\end{align}
which can be regarded as a new message passing equation for the derivative~$\eta_{y}^{i\leftarrow j}$.  To apply it, we first solve for the $q^{i\leftarrow j}_{y}(\beta)$ in the usual fashion then fix their values and iterate~\eqref{dmessage} from a suitable initial condition until convergence.

For large neighborhoods, where the sums over spins states cannot be performed exhaustively, the local Monte Carlo procedure described in the main text carries over naturally.  We define
\begin{equation}
\label{ijshort}
\langle A\rangle_{N_{j\setminus i}}\! = \!\sum_{\vec{s}_{N_{j\setminus i}}} \!A(\vec{s}_{N_{j\setminus i}})\frac{e^{-\beta H_{N_{j\setminus i}}(\vec{s}_{N_{j\setminus i}})} \!\!\prod\limits_{k\in N_{j\setminus i}\setminus j}q^{j\leftarrow k}_{s_k}(\beta)}{Z^{i\leftarrow j}(\beta)}
\end{equation}
and then rewrite Eq.~\eqref{dmessage} as an average
\begin{align}
\eta_{y}^{i\leftarrow j} &= \biggl\langle\left[q^{i\leftarrow j}_{y}(\beta)-\delta_{s_j,y}\right]H_{N_{j\setminus i}}(\vec{s}_{N_{j\setminus i}})\nonumber \\
&\qquad{}+ \left[\delta_{s_j,y}-q^{i\leftarrow j}_{y}(\beta)\right]\sum\limits_{k\in N_{j\setminus i}\setminus j}\frac{\eta^{j\leftarrow k}_{s_k}(\beta)}{q^{j\leftarrow k}_{s_k}(\beta)}\biggr\rangle_{\!\!\!N_{j\setminus i}},
\end{align}
which can be evaluated using Monte Carlo sampling as previously.

We can also differentiate $Z^i(\beta)$, Eq.~\eqref{Zs}, which yields
\begin{align}
&\frac{1}{Z^{i}(\beta)}\frac{dZ^{i}(\beta)}{d\beta} =\frac{1}{Z^{i}(\beta)}\sum\limits_{\vec{s}_{N_i}}e^{-\beta H_{N_i}(\vec{s}_{N_i})}\prod\limits_{j\in N_i\setminus i}q^{i\leftarrow j}_{s_j}(\beta)  \nonumber\\
&\qquad\times\biggl[ \sum\limits_{j\in N_i\setminus i}\frac{1}{q^{i\leftarrow j}_{s_j}(\beta)}\frac{dq^{i\leftarrow j}_{s_j}(\beta)}{d\beta}- H_{N_i}(\vec{s}_{N_i}) \biggr],
\end{align}
which can again be written as an average
\begin{equation}
\frac{1}{Z^{i}(\beta)}\frac{dZ^{i}(\beta)}{d\beta}
 =\biggl\langle \sum\limits_{j\in N_i\setminus i}\frac{\eta^{i\leftarrow j}_{s_j}}{q^{i\leftarrow j}_{s_j}(\beta)} - H_{N_i}(\vec{s}_{N_i})\biggr\rangle_{\!\!N_i},
\label{dZ}
\end{equation}
where we have used a shorthand analogous to that of Eq.~\eqref{ijshort}:
\begin{align}
\langle A\rangle_{N_i} = \sum\limits_{\vec{s}_{N_i}}A(\vec{s}_{N_i})\frac{e^{-\beta H_{N_i}(\vec{s}_{N_i})}\prod\limits_{j\in N_i\setminus i}q^{i\leftarrow j}_{s_j}(\beta) }{Z^{i}(\beta)} .
\end{align}

\begin{widetext}
Differentiating Eq.~\eqref{app:U_MP} and substituting from Eqs.~\eqref{dmessage} and~\eqref{dZ} we now find, after some manipulation, that
\begin{align}
\label{final_dudbeta}
\frac{dU}{d\beta} &= \frac12 \sum_{i\in V}\bigl[\bigl\langle H_{\partial_i}(\vec{s}_{\partial_i}) \bigr\rangle_{N_i}\bigl\langle H_{N_i}(\vec{s}_{N_i})\bigr\rangle_{N_i}-\bigl\langle H_{\partial_i}(\vec{s}_{\partial_i}) H_{N_i}(\vec{s}_{N_i})\bigr\rangle_{N_i}\bigr]\nonumber\\
&\hspace{4em}{} + \frac12 \sum_{i\in V}\Biggl[\Biggl\langle H_{\partial_i}(\vec{s}_{\partial_i}) \sum_{j\in N_i\setminus i}\frac{\eta^{i\leftarrow j}_{s_j}}{q^{i\leftarrow j}_{s_j}}\Biggr\rangle_{\!\!N_i}-\bigl\langle H_{\partial_i}(\vec{s}_{\partial_i}) \bigr\rangle_{N_i} \Biggl\langle \sum_{j\in N_i\setminus i}\frac{\eta^{i\leftarrow j}_{s_j}}{q^{i\leftarrow j}_{s_j}}\Biggr\rangle_{\!\!N_i} \,\Biggr],
\end{align}
which can be substituted into Eq.~\eqref{C_def} to calculate~$C$.

\subsection{Local Monte Carlo simulation for the Ising model}
\label{SI:local_mcmc}
As discussed in the main text, when neighborhoods are too large to allow us to sum exhaustively over their states we can approximate the message passing equations by Monte Carlo sampling.  Taking again the example of the Ising model, the message passing equations are
\begin{equation}
q^{i} = \frac{\sum_{\vec{s}_{N_i}}\delta_{s_i,+1}e^{-\beta H_{N_i}(\vec{s}_{N_i})}\prod_{j\in N_i\setminus i}q^{i\leftarrow j}_{s_j}}{\sum_{\vec{s}_{N_i}}e^{-\beta H_{N_i}(\vec{s}_{N_i})}\prod_{j\in N_i\setminus i}q^{i\leftarrow j}_{s_j} }, \qquad
q^{i\leftarrow j} = \frac{\sum_{\vec{s}_{N_{j\setminus i}}}\delta_{s_j,+1}e^{-\beta H_{N_{j\setminus i}}(\vec{s}_{N_{j\setminus i}})}\prod_{k\in N_{j\setminus i}\setminus j}q^{j\leftarrow k}_{s_k}}{\sum_{\vec{s}_{N_{j\setminus i}}}e^{-\beta H_{N_{j\setminus i}}(\vec{s}_{N_{j\setminus i}})}\prod_{k\in N_{j\setminus i}\setminus j}q^{j\leftarrow k}_{s_k} },
\end{equation}
where the messages in this case represent the probability of the corresponding spin being~$+1$.  If we divide top and bottom by $\sum_{\vec{s}_{N_i}}e^{-\beta H_{N_i}(\vec{s}_{N_i})}$ in the first equation and by $\sum_{\vec{s}_{N_{j\setminus i}}}e^{-\beta H_{N_{j\setminus i}}(\vec{s}_{N_{j\setminus i}})}$ in the second, we get
\begin{align}
q^{i} &= \frac{\sum_{\vec{s}_{N_i}}e^{-\beta H_{N_i}(\vec{s}_{N_i})}\bigl(\delta_{s_i,+1} \prod_{j\in N_i\setminus i}q^{i\leftarrow j}_{s_j}\bigr)\big/\sum_{\vec{s}_{N_i}}e^{-\beta H_{N_i}(\vec{s}_{N_i})}}{\sum_{\vec{s}_{N_i}}e^{-\beta H_{N_i}(\vec{s}_{N_i})}\bigl(\prod_{j\in N_i\setminus i}q^{i\leftarrow j}_{s_j}\bigr) \big/ \sum_{\vec{s}_{N_i}}e^{-\beta H_{N_i}(\vec{s}_{N_i})}}, \\
q^{i\leftarrow j} &= \frac{\sum_{\vec{s}_{N_{j\setminus i}}}e^{-\beta H_{N_{j\setminus i}}(\vec{s}_{N_{j\setminus i}})}\bigl( \delta_{s_j,+1} \prod_{k\in N_{j\setminus i}\setminus j}q^{j\leftarrow k}_{s_k}  \bigr)
\big/ \sum_{\vec{s}_{N_{j\setminus i}}}e^{-\beta H_{N_{j\setminus i}}(\vec{s}_{N_{j\setminus i}})}}{\sum_{\vec{s}_{N_{j\setminus i}}}e^{-\beta H_{N_{j\setminus i}}(\vec{s}_{N_{j\setminus i}})}\bigl( \prod_{k\in N_{j\setminus i}\setminus j}q^{j\leftarrow k}_{s_k}  \bigr)
\big/ \sum_{\vec{s}_{N_{j\setminus i}}}e^{-\beta H_{N_{j\setminus i}}(\vec{s}_{N_{j\setminus i}})}}.
\end{align}
\end{widetext}

Numerators and denominators now take the form of a Boltzmann average, but over the distributions defined by $H_{N_i}$ and $H_{N_{j\setminus i}}$ alone, which we can think of as a ``zero-field'' ensemble that omits the effect of the ``external field'' imposed by the messages.  Defining the useful shorthand
\begin{align}
\label{zeroshort}
\av{A}_{0,N_i} &= \frac{\sum_{\vec{s}_{N_i}}e^{-\beta H_{N_i}(\vec{s}_{N_i})}A(\vec{s}_{N_{i}})}{\sum_{\vec{s}_{N_i}}e^{-\beta H_{N_i}(\vec{s}_{N_i})}}, \\
\av{A}_{0,N_{j\setminus i}}&=\frac{\sum_{\vec{s}_{N_{j\setminus i}}}e^{-\beta H_{N_{j\setminus i}}(\vec{s}_{N_{j\setminus i}})}A(\vec{s}_{N_{j\setminus i}})}{\sum_{\vec{s}_{N_{j\setminus i}}}e^{-\beta H_{N_{j\setminus i}}(\vec{s}_{N_{j\setminus i}})}},
\end{align}
we can then write the message passing equations in the form
\begin{align}
q^{i} &= \frac{\bigl\langle \delta_{s_i,+1} \prod\limits_{j\in N_i\setminus i}q^{i\leftarrow j}_{s_j} \bigr\rangle_{0,N_i}}{\bigl\langle \prod\limits_{j\in N_i\setminus i}q^{i\leftarrow j}_{s_j} \bigr\rangle_{0,N_i}}, \\
q^{i\leftarrow j} &= \frac{\bigl\langle \delta_{s_j,+1} \prod\limits_{k\in N_{j\setminus i}\setminus j}q^{j\leftarrow k}_{s_k}  \bigr\rangle_{0,N_{j\setminus i}}}{\bigl\langle \prod\limits_{k\in N_{j\setminus i}\setminus j}q^{j\leftarrow k}_{s_k}  \bigr\rangle_{0,N_{j\setminus i}}},
\end{align}
where the ``$0$'' serves to remind us that the expectation is over the zero-field ensemble.  Expressing the equations as zero-field expectations allows us to evaluate them using the Wolff algorithm, which is highly efficient in this context.

We can further speed up sampling by making use of the up-down symmetry of the zero-field ensemble, which effectively gives us two samples for every spin state.  If we obtain a set of samples $\{\vec{s}_N\}$ by sampling from the zero-field ensemble, then because of symmetry $\{-\vec{s}_N\}$ are also correct samples that would have occurred with the same probability.  Including these additional samples explicitly in the message passing equations gives
\begin{align}
&q^{i\leftarrow j} = \nonumber\\
&\frac{\bigl\langle \delta_{s_j,+1} \prod\limits_{k\in N_{j\setminus i}\setminus j}q^{j\leftarrow k}_{s_k}+\delta_{-s_j,+1} \prod\limits_{k\in N_{j\setminus i}\setminus j}(1-q^{j\leftarrow k}_{s_k})  \bigr\rangle_{0,N_{j\setminus i}}}{\bigl\langle \prod\limits_{k\in N_{j\setminus i}\setminus j}q^{j\leftarrow k}_{s_k} + \prod\limits_{k\in N_{j\setminus i}\setminus j}(1-q^{j\leftarrow k}_{s_k})  \bigr\rangle_{0,N_{j\setminus i}}},
\end{align}
and corresponding expressions can be derived for any expectation.

\subsection{The Jacobian at the critical point}
\label{SI:jacobian}
In the main text we used the leading eigenvalue of the Jacobian of the message passing iteration at the trivial fixed point to locate the position of the phase transition.  Taking the Ising model as our example once again, the calculation is as follows.

The message passing equations can be rewritten as
\begin{align}
q^{i\leftarrow j} &= \frac{1}{Z^{i\leftarrow j}}\sum\limits_{\vec{s}_{N_{j\setminus i}}} \tfrac12(1+s_j)\,e^{-\beta H_{N_{j\setminus i}}(\vec{s}_{N_{j\setminus i}})}
\nonumber\\
&\qquad{} \times\prod\limits_{k\in N_{j\setminus i}\setminus j}\bigl[\tfrac12(1-s_k)+s_kq^{j\leftarrow k}\bigr],
\end{align}
where
\begin{align}
Z^{i\leftarrow j} = \sum\limits_{\vec{s}_{N_{j\setminus i}}}e^{-\beta H_{N_{j\setminus i}}(\vec{s}_{N_{j\setminus i}})} \!\!\!\!\prod_{k\in N_{j\setminus i}\setminus j} \!\!\bigl[\tfrac12(1-s_k) + s_kq^{j\leftarrow k}\bigr].
\end{align}
Considering the sum over spins as a local average again, the elements of the Jacobian are then given by
\begin{align}
\frac{\partial q^{i\leftarrow j}}{\partial q^{\mu\leftarrow \nu}} &= \mathbf{1}_{\{\mu=j,\nu\in N_{j\setminus i}\}}\biggl[\biggl\langle  \frac{(1+s_j)s_\nu}{1-s_\nu+2s_\nu q^{\mu\leftarrow \nu}}\biggr\rangle_{\!N_{j\setminus i}}\nonumber \\
&\qquad- \bigl\langle1+s_j\bigr\rangle_{N_{j\setminus i}}
\biggl\langle\frac{s_\nu}{1-s_\nu+2s_\nu q^{\mu\leftarrow \nu}}\biggr\rangle_{\!N_{j\setminus i}} \biggr],
\end{align}
where $\mathbf{1}_{ \{ \dots \} }$ is the indicator function and we have used the shorthand from Eq.~\eqref{ijshort} again.  Now evaluating this expression at the trivial fixed point $q^{j\leftarrow k} = \frac12$ for all $j,k$ (which we write as simply $q=\frac12$ for short), we get the Jacobian
\begin{align}
J_{j\to i,\nu\to\mu} = \frac{\partial q^{i\leftarrow j}}{\partial q^{\mu\leftarrow \nu}}\biggr\vert_{q=\frac12} = \tilde B_{j\to i,\nu\to\mu}D_{j\to i,\nu\to\mu},
\end{align}
where $\tilde B$ is a generalization of the non-backtracking matrix given by
\begin{align}
\tilde B_{j\to i,\nu\to\mu} = \biggl\lbrace\begin{array}{ll}
1 & \qquad \mbox{if $\mu=j$ and $\nu\in N_{j\setminus i}$,} \\
0 & \qquad \mbox{otherwise,}
\end{array}
\end{align}
and
\begin{align}
&D_{j\to i,\nu\to\mu} =  \frac{\sum\limits_{\vec{s}_{N_{j\setminus i}}} s_\mu s_\nu \,e^{-\beta H_{N_{j\setminus i}}(\vec{s}_{N_{j\setminus i}})}}{\sum\limits_{\vec{s}_{N_{j\setminus i}}}e^{-\beta H_{N_{j\setminus i}}(\vec{s}_{N_{j\setminus i}})}} \\
&- \frac{\sum\limits_{\vec{s}_{N_{j\setminus i}}} s_\mu\,e^{-\beta H_{N_{j\setminus i}}(\vec{s}_{N_{j\setminus i}})}}{\sum\limits_{\vec{s}_{N_{j\setminus i}}}e^{-\beta H_{N_{j\setminus i}}(\vec{s}_{N_{j\setminus i}})}}\times \frac{\sum\limits_{\vec{s}_{N_{j\setminus i}}} s_\nu\,e^{-\beta H_{N_{j\setminus i}}(\vec{s}_{N_{j\setminus i}})}}{\sum\limits_{\vec{s}_{N_{j\setminus i}}}e^{-\beta H_{N_{j\setminus i}}(\vec{s}_{N_{j\setminus i}})}},\nonumber
\end{align}
which we note is temperature dependent.  Using the shorthand from Eq.~\eqref{zeroshort}, $D$~can also be written in the simpler form
\begin{align}
D_{j\to i,\nu\to\mu} = \av{s_\mu s_\nu}_{0,N_{j\setminus i}} - \av{s_\mu }_{0,N_{j\setminus i}} \av{s_\nu}_{0,N_{j\setminus i}} .
\label{betterd}
\end{align}

At the temperature where the magnitude of the leading eigenvalue $\lambda_\textrm{max}$ of $J$ is 1 at the trivial fixed point, the fixed point transitions from being stable to unstable, which corresponds to the phase transition as described in the main text.  Thus we can locate the phase transition by evaluating the matrices $\tilde B$ and $D$ numerically and using them to compute $\abs{\lambda_\textrm{max}}$.  Note that the expectations in Eq.~\eqref{betterd} do not depend on the values of the messages, so we do not need to perform message passing to calculate them---evaluating the Jacobian and locating the phase transition requires us only to perform the sums over neighborhoods or approximate them using local Monte Carlo.

\subsection{Proof of neighborhood-level factorization}
\label{SI:factorization}
In the calculation of the partition function and entropy in Section~\ref{main:neig_factorization} of the main text we make use of the factorized form
\begin{align}
P(\vec{s}) = { \prod\limits_{i \in G} P(\vec{s}_{N_{i}}) \over \prod\limits_{((i,j)) \in G} P(\vec{s}_{\cap_{ij}})^{2 / \vert \cap_{ij} \vert}}\,,
\label{app:neighborhood-ansatz-1}
\end{align}
where $\cap_{ij} = {N}_{i} \cap {N}_{j}$ and $((i,j))$ are pairs of nodes that are contained in each other's neighborhood, i.e.,~nodes~$i$ and $j$ such that $i \in {N}_{j}$ and $j \in {N}_{i}$.  This form is derived as follows.

\begin{figure*}
\centering
\includegraphics[width=1.0\linewidth]{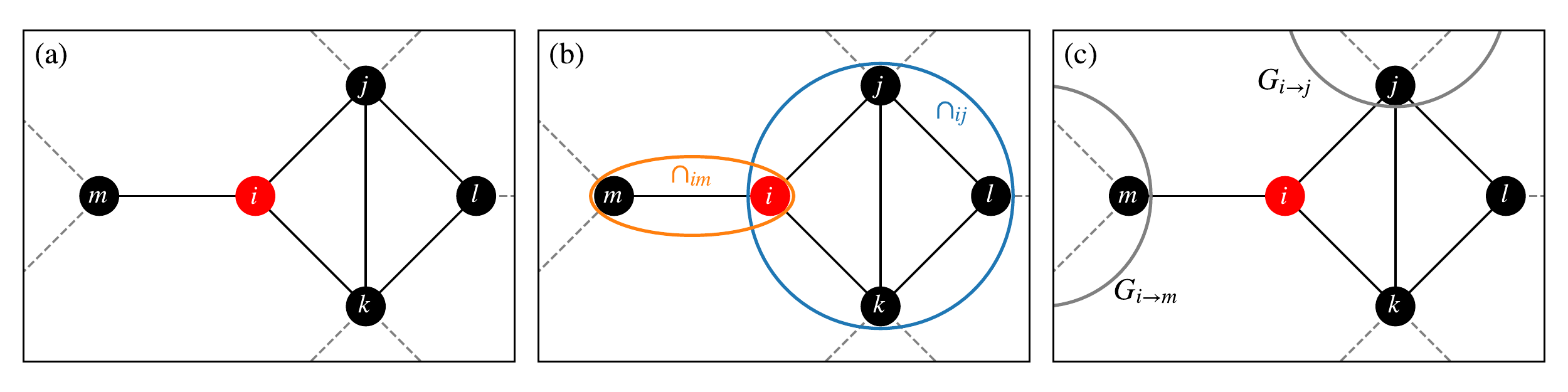}
\caption[Neighborhoods and various related quantities in an example network.]{\label{fig:neighborhoods}\textbf{Neighborhoods and various related quantities for a node~$i$ in an example network.}  In this example we assume that $r=2$ is sufficient to capture all primitive cycles and thus that calculations at $r=2$ are exact.  (a)~The neighborhood ${N}_i = {N}_i^{(2)}$ contains the edges and nodes shown in solid black.  (b)~At node~$i$ there are two distinct intersections, $\cap_{im} = {N}_i \cap {N}_m$ and $\cap_{ij} = {N}_i \cap {N}_j$.  Note that the intersections for all pairs of nodes in $\cap_{ij}$ are identical.  For instance in this example we have $\cap_{ij} = \cap_{ik} = \cap_{il} = \cap_{jk} = \cap_{jl} = \cap_{lk}$.  (c)~The subgraph $G_{i \to j}$ is the connected component to which~$j$ belongs after all edges in ${N}_i$ are removed, and similarly for~$G_{i \to m}$.}
\end{figure*}

Consider Fig.~\ref{fig:neighborhoods}, which illustrates the definition of the sets of nodes we use and their intersections.  As shown in panel~(b) of the figure, many of the sets are equivalent to one another.  Specifically, for any pair $k,l \in \cap_{ij}$ we have $\cap_{kl} = \cap_{ij}$.  This allows us to write
\begin{equation}
P(\vec{s}_{\cap_{ij}}) = \biggl[ \prod\limits_{(k,l) \in \cap_{ij}}
\!\!\!\!P(\vec{s}_{\cap_{kl}}) \>\biggr]^{1 / {\vert \cap_{ij} \vert \choose 2} }\!\!\!
  = \!\!\!\!\prod\limits_{(k,l) \in \cap_{ij}} \!\!\!\!
    P(\vec{s}_{\cap_{kl}})^{1 / {\vert \cap_{kl} \vert \choose 2} },
\label{intersection-equiv}
\end{equation}
where the product is over all ${\vert \cap_{ij} \vert \choose 2}$ pairs $\{k,l\} \in \cap_{ij}$.  A~proof of Eq.~\eqref{app:neighborhood-ansatz-1} can then be achieved by induction.  Assume that the formula is correct for all networks with fewer than $n$ nodes and no primitive cycles longer than $r+2$.
If $G$ is a network with $n$ nodes and no primitive cycles longer than $r+2$ then
\begin{align}
P(\vec{s}) &= P(\vec{s}_{{N}_i}) \prod\limits_{j \in {N}_{i}} P( \vec{s}_{{N}_j} \vert \vec{s}_{{N}_i}) P(\vec{s}_{G_{i \to j}} \vert \vec{s}_{{N}_j}) \nonumber\\
  &= P(\vec{s}_{{N}_i}) \prod\limits_{j \in {N}_{i}} {P( \vec{s}_{{N}_j}) \over P( \vec{s}_{\cap_{ij}})} P(\vec{s}_{G_{i \to j}} \vert \vec{s}_{{N}_j \setminus {N}_i}),
\end{align}
where $G_{i \to j}$ denotes the connected subgraph to which~$j$ belongs after all edges in~$N_i$ have been removed (see Fig.~\ref{fig:neighborhoods}).
Since by definition the $G_{i \to j}$ have fewer than $n$ nodes and no primitive cycles longer than~$r+2$, Eq.~\eqref{app:neighborhood-ansatz-1} is by hypothesis true for these subgraphs, and using~\eqref{intersection-equiv} we have
\begin{align}
P(\vec{s}) &= P(\vec{s}_{{N}_i}) \prod\limits_{j \in {N}_i}  \frac{1}{\prod\limits_{(k,l) \in \cap_{ij}} P(\vec{s}_{\cap_{kl}})^{1 / {\vert \cap_{kl} \vert \choose 2} } } \nonumber\\
&\hspace{8em}{} \times { \prod\limits_{k \in G_{i \to j}} P(\vec{s}_{{N}_k}) \over \prod\limits_{(k,l) \in G_{i \to j}} P(\vec{s}_{\cap_{kl}})^{2 / \vert \cap_{kl} \vert}} \nonumber\\
 &= { \prod\limits_{i \in G} P(\vec{s}_{{N}_i}) \over \prod\limits_{(i,j) \in G} P(\vec{s}_{\cap_{ij}})^{2 / \vert \cap_{ij} \vert}}.
\end{align}
The base case is a graph with a single node, for which \eqref{app:neighborhood-ansatz-1} is trivially true, and hence by induction~\eqref{app:neighborhood-ansatz-1} is true for all networks that have no primitive cycles longer than~$r+2$.

For the purposes of the calculation presented in Section~\ref{main:neig_factorization}, Eq.~\eqref{app:neighborhood-ansatz-1} can be further simplified by noting that
\begin{align}
P(\vec{s}_{{N}_i}) &= P(s_i) \prod\limits_{j \in {N}_i} P( \vec{s}_{\cap_{ij}} \vert s_i )^{1 \over \vert \cap_{ij} \vert -1}\nonumber\\
  &= P(s_i) \prod\limits_{j \in {N}_i} \biggl[ {P( \vec{s}_{\cap_{ij}}  ) \over P(s_i) } \biggr]^{1 \over \vert \cap_{ij} \vert -1}.
\label{P_neighborhood}
\end{align}
Substituting this result into~\eqref{app:neighborhood-ansatz-1} then yields
\begin{align}
P(\vec{s}) = \prod\limits_{((i,j)) \in G} \!\!\! P( \vec{s}_{\cap_{ij}})^{1 / {\vert \cap_{ij} \vert \choose 2}}
  \prod\limits_{(i,j) \in G} \!\!P(s_i,s_j)^{W_{ij}} \nonumber \\
  \times\prod\limits_{i \in G} \,P(s_i)^{C_i},
\label{app:N_factorization}
\end{align}
where
\begin{equation}
W_{ij} = 1 - \!\!\!\sum\limits_{ ((l,m)) \in G } { 1 \over { \vert \cap_{lm} \vert \choose 2 }} \mathbf{1}_{ \{ (i,j) \in \cap_{lm} \} }
\end{equation}
and
\begin{equation}
C_i = 1 - \sum\limits_{j \in {N}_i} {1 \over \vert \cap_{ij} \vert -1}
  - \sum\limits_{j \in N_i^{(0)}} W_{ij}.
\end{equation}
The one- and two-spin marginals $P(s_i)$ and $P(s_i,s_j)$ can be calculated using the message passing methods described in the text, while the intersection marginal~$P( \vec{s}_{\cap_{ij}})$ is given by
\begin{equation}
P(\vec{s}_{\cap_{ij}}) = \frac{1}{Z_{\cap_{ij}}} e^{-\beta H(\vec{s}_{\cap_{ij}})} q_{i
\leftarrow j}(s_j) \prod\limits_{k \in \cap_{ij} \setminus j } q_{j \leftarrow k} (s_k),
\end{equation}
where $H(\vec{s}_{\cap_{ij}})$ denotes the terms of the full Hamiltonian that fall in~$\cap_{ij}$ and $Z_{\cap_{ij}}$ is the corresponding normalizing constant.

Equation~\eqref{app:N_factorization} is exact when the network contains no primitive cycles longer than~$r+2$, in which case $W_{ij}=0$.  When there are longer primitive cycles (and hence Eq.~\eqref{app:neighborhood-ansatz-1} is not exact), the terms $P(s_i,s_j)^{W_{ij}}$ ensure that each edge gets weighted correctly in the factorization.

\end{document}